\documentclass[aps,prb,twocolumn,groupedaddress,showpacs,preprintnumbers,amsmath
,amssymb]{revtex4}
\usepackage{graphicx}

\def\tE{\tau_{\mathrm{E}}}
\def\tD{\tau_{\mathrm{D}}}
\def\TH{T_{\mathrm{H}}}
\def\pF{p_{\mathrm{F}}}
\def\tW{\tau_{\mathrm{W}}}

\newcommand{\eref}[1]{(\ref{#1})}
\newcommand{\erefs}[2]{(\ref{#1},\ref{#2})}
\newcommand{\erefss}[3]{(\ref{#1},\ref{#2},\ref{#3})}

\newcommand{\ocite}[1]{Ref.\ \onlinecite{#1}}
\newcommand{\ocites}[1]{Refs.\ \onlinecite{#1}}

\begin{document}

\title{Ehrenfest-time Dependence of Counting Statistics for Chaotic Ballistic
Systems}

\author{Daniel Waltner, Jack Kuipers, and Klaus Richter}
\affiliation{Institut f\"ur Theoretische Physik, Universit\"at Regensburg,
D-93040 Regensburg, Germany}

\date{\today}

\begin{abstract}
Transport properties of open chaotic ballistic systems and their statistics can
be expressed in terms of the scattering matrix connecting incoming and outgoing
wavefunctions.  Here we calculate the dependence of correlation functions of
arbitrarily many pairs of scattering matrices at different energies on the
Ehrenfest time using trajectory based semiclassical methods.  This enables us to
verify the prediction from effective random matrix theory that one part of the
correlation function obtains an exponential damping depending on the Ehrenfest
time, while also allowing us to obtain the additional contribution that arises
from bands of always correlated trajectories.  The resulting Ehrenfest-time
dependence, responsible e.g.\ for secondary gaps in the density of states of
Andreev billiards, can also be seen to have strong effects on other transport
quantities such as the distribution of delay times.
\end{abstract}
\maketitle

\section{Introduction}

After the conjecture by Bohigas, Gianonni and Schmit in 1984 \cite{Boh}, that
chaotic systems are well described by random matrix theory (RMT) \cite{Met},
research started to demonstrate this connection on dynamical grounds by means of
semiclassical methods based on analyzing energy averaged products of expressions
similar to the Gutzwiller trace formula \cite{Gut} for the density of states,
that are asymptotically exact in the limit $\hbar\rightarrow 0$. For open
systems we are particularly interested in the scattering matrix $S(E)$, which is
an $N\times N$ matrix if the scattering leads carry $N$ states or channels in
total.  Its elements can, like the Gutzwiller trace formula, be expressed
\cite{Richter} in terms of sums over the classical trajectories containing the
stability factors of the orbits $A_\gamma$ and rapidly oscillating phases
depending on the classical actions $S_\gamma$ of the considered trajectories
$\gamma$ divided by $\hbar$
\begin{equation}
\label{eq1}
S_{o,i}\approx\frac{1}{\sqrt{\TH}}\sum_{\gamma(i\to o)} A_\gamma{\rm e}^{({\rm
i}/\hbar) S_\gamma},
\end{equation}
with $\TH\equiv2\pi\hbar \Delta$ with the mean level spacing of the quantum
system $\Delta$. Here the sum is over the scattering trajectories that connect
the two channels $i$ and $o$.  For systems with two (or more) leads the
scattering matrix breaks up into reflecting and transmitting subblocks, so we
might restrict our attention to trajectories starting and ending in certain
leads.

In the context of spectral statistics, i.e.\ for the two point correlation
function of the density of states containing a double sum over periodic orbits,
this dynamical understanding of the conjecture \cite{Boh} was - as for other
quantities - achieved in several steps.  Starting with the pairing of identical
(or time reversed orbits in the presence of time reversal symmetry) the so
called diagonal contribution was evaluated in \ocite{Ber1} using a sum rule from
\ocite{Han}. Nondiagonal contributions consisting of pairs of long orbits
differing essentially only in the place where one of the orbits possesses a self
crossing and the other avoids this crossing were analyzed in \ocite{Sie}. This
was extended \cite{Mul1} and formalized for orbits differing at several places,
so called encounters.

In the context of transport, i.e.\ for example for the two-point correlator of
scattering matrix elements, which if restricted to the transmission subblocks is
via the Landauer-B\"uttiker formalism \cite{Lan} proportional to the
conductance, the diagonal contribution was calculated in \ocite{Jal}. An orbit
pair differing only in one crossing was analyzed in \ocite{Ric} and this was
again extended to orbits differing at several places\cite{Heu1}. These results
and those for closed systems agreed with results from RMT, but besides this
dynamical understanding of the RMT results, these semiclassical calculations
proved very successful in determining the effect of a finite Ehrenfest time
$\tE$ on transport quantities, starting with the pioneering work of \ocite{Ada}.
The Ehrenfest time \cite{Chi} separates times when the time evolution of a
particle follows essentially the classical dynamics from times when it is
dominated by wave interference. Its value is obtained as the time when two
points inside a wave packet initially of quantum size $\hbar/\pF$ with the Fermi
momentum $\pF$ evolve to points with a distance $L$ of the linear system size.
We thus get due to the exponential separation of neighboring trajectories in the
chaotic case
\begin{equation}
\tE=\frac{1}{\lambda}\ln\frac{\pF L}{\hbar} ,
\end{equation}
with the Lyapunov exponent $\lambda$.

Before these semiclassical calculations of the Ehrenfest-time dependence,
there already existed theories to describe the effect of a finite Ehrenfest time on the
correlators of scattering matrix elements: Aleiner and Larkin obtained
\cite{Aleiner} for the correlator of two transmission matrices, i.e.\ the
conductance, an exponential suppression with increasing Ehrenfest time in
agreement with semiclassics.  This work was however unsatisfactory in one main
aspect: a small amount of impurity scattering was introduced by hand to imitate
the effects of diffraction in a ballistic system.

Another phenomenological theory to describe the effect of a finite Ehrenfest
time is effective RMT \cite{Sil}.  It splits the phase space
and thereby also the underlying scattering matrix of the considered system into
a classical and a quantum part, where the first one is determined by all
trajectories shorter than $\tE$ and the second one by all trajectories longer
than $\tE$, as well as introducing an artificial phase dependent on the
Ehrenfest time. The predictions of this theory are only partially correct: weak
localization is predicted to be independent of the Ehrenfest time, while the
previously mentioned theories and also numerical simulations
\cite{Rahav,Jacquod} predict it to decay with the Ehrenfest time. In contrast to
the quantum correction of weak localization, effective RMT gave good predictions
for effects at leading order in $N$ such as shot noise
\cite{Agam,Silvestr,Tworzyd,Whi} or the gap in the density of states of a
chaotic Andreev billiard \cite{Goorden,Ben}.

Staying only at the leading order in inverse channel number we will consider the
correlation function of $2n$ scattering matrices at alternating energies defined
as
\begin{equation}
C(\epsilon,n,\tau) =\frac{1}{N}{\rm
Tr}\left[S^{\dagger}\left(-\frac{\epsilon\hbar}{2\tD}\right)S\left(+\frac{
\epsilon\hbar}{2\tD}\right)\right]^n  , 
\end{equation}
where for simplicity the energy $\epsilon$ is measured with respect to the
(Fermi) energy $E$ and in units of the so called Thouless energy $E_{\rm T} =
\hbar /2\tD$ with the dwell time $\tD$ measuring the typical time a particle
stays inside the system. The latter is related to the Heisenberg time $\TH$ via
the relation $\TH=N\tD$. The Ehrenfest-time dependence is incorporated in
$\tau\equiv\tE/\tD$.  The explicit form is 
\begin{eqnarray}
\label{eq4}
C(\epsilon,\tau,n) &=& C_1(\epsilon,\tau,n)+C_2(\epsilon,\tau,n), \\
\label{eq5}
C_{1}(\epsilon,\tau,n) & = & C(\epsilon,n){\rm e}^{-\tau(1-{\rm i}n\epsilon)},
\\
\label{eq6}
C_{2}(\epsilon,\tau,n) & = & \frac{1-{\rm e}^{-\tau(1-{\rm i}n\epsilon)}}{1-{\rm
i}n\epsilon},
\end{eqnarray}
with the RMT (i.e.\ $\tau=0$) part of this correlation function denoted by
$C(\epsilon,n)$. The term in (\ref{eq5}) derives from effective RMT
\cite{Sil,Bro}.  Although this theory describes certain phenomena quite well,
e.g.\ the dependence of the Andreev gap on the Ehrenfest time \cite{Ben}, a
dynamical justification of this result is still lacking.  So far Ref.
\onlinecite{Bro} calculated $C(\epsilon,\tau,n)$ for $n=1,2,3$ while
\ocites{Jac,Jacquod} showed the separation into two terms in \eref{eq4} to be a
consequence of the preservation under time evolution of a phase-space volume of
the system.  Moreover they also calculated the explicit form we give in
\eref{eq6} for the second term and that the first term in \eref{eq5} is
proportional to the factor ${\rm e}^{-\tau(1-{\rm i}n\epsilon)}$.

\begin{figure}
\begin{center}
\includegraphics[width=\columnwidth]{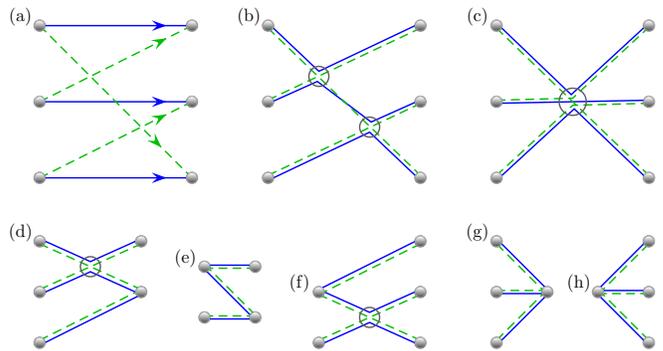}
\caption{(Color online) The trajectory sets with encounters that contribute to
the 3rd correlation function $C(\epsilon,3)$}
\label{fig1}
\end{center}
\end{figure}

Because of \eref{eq1} the correlation function can be written semiclassically in
terms of $2n$ scattering trajectories connecting channels along a closed cycle
like in Fig.\ \ref{fig1}a.  This leads to trajectory sets with encounters as in
Fig.\ \ref{fig1}b,c which can then be moved into the leads to create the
remaining diagrams in Fig.\ \ref{fig1}.  Including the correct prefactors and
the energy dependence, the correlation function becomes semiclassically
\begin{eqnarray}
\label{eq6a}
C(\epsilon,\tau,n)  & \approx &  \frac{1}{N {\TH}^{n}}\prod_{j=1}^{n}
\sum_{i_{j},o_{j}} 
 \sum_{\substack{\gamma_{j}(i_{j}\to o_{j}) \\ \gamma_{j}' (i_{j+1}\to o_{j})}}
A_{\gamma_{j}}A_{\gamma_{j}'}^{*} \nonumber\\
&& \times {\rm e}^{({\rm i}/\hbar)(S_{\gamma_{j}}-S_{\gamma_{j}'})}
  {\rm e}^{({\rm i}\epsilon/2) (T_{\gamma_{j}}+T_{\gamma_{j}'})/\tD}, 
\end{eqnarray}
$T_{\gamma}$ are the times trajectories $\gamma$ spend inside the system, and we
identify the channels $i_{n+1}=i_{1}$. Note that \eref{eq6a} and this
identification imply that the trajectories and their partners (traversed in
reversed direction) considered in $C(\epsilon,\tau,n)$ form a closed cycle.

In this paper we want to show how \erefss{eq4}{eq5}{eq6} can be obtained using
the trajectory based methods developed in \ocites{Sie,Mul1,Ric,Heu1}. In Sec.\
\ref{part1} we consider the first term in \eref{eq4}: we show that the prefactor
$C(\epsilon,n)$ of the exponential is indeed given by the RMT expression
obtained in \ocite{Kui} and that this is multiplied by the exponential given in
\eref{eq5}. The underlying diagrams considered here are the same as the ones
occurring also in the semiclassical calculation of the RMT contribution. In
Sec.\ \ref{part2} we consider the second term in \eref{eq4} and show how this
contribution arises from trajectories that are always correlated. Furthermore we
show in Sec.\ \ref{mixing} that there exist no mixed terms between the first
and the second term in \eref{eq4}, that could result - expressed in terms of the
considered diagrams - from correlations between trajectories always correlated
with each other on the one side and trajectories only correlated with each other
during encounters on the other side. 

\section{Influence of the Ehrenfest time on trajectories with
encounters}\label{part1}

\begin{figure}
\begin{center}
\includegraphics[width=0.8\columnwidth]{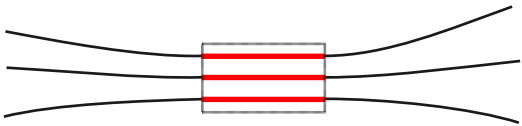}
\caption{(Color online) A 3-encounter as it can be approximated in the
RMT-treatment (c.f.\ Fig.\ \ref{fig1}c). The encounter stretches are marked by a
box (shown red).}
\label{fig2}
\end{center}
\end{figure}

\begin{figure}
\begin{center}
\includegraphics[width=0.8\columnwidth]{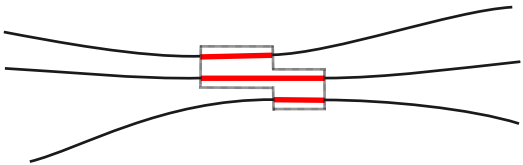}
\caption{(Color online) A 3-encounter as previously treated with Ehrenfest time
\cite{Bro}. The encounter stretches are marked by a box (shown red).}
\label{fig3}
\end{center}
\end{figure}

The main idea in this section is to split our diagrams in a different way
compared to the semiclassical analysis without Ehrenfest time (referred to as
the RMT-treatment) and the analysis of the Ehrenfest-time dependence of the
cases $n=1,2,3$ in \ocite{Bro}: in the semiclassical calculation one considers
an arbitrary number of orbits encountering each other. It turns out in the
RMT-treatment to be sufficient to consider only encounters where all orbits are
linearizable up to the {\it same} point, see for example Fig.\ \ref{fig2}.
When taking into account the Ehrenfest-time dependence this is no longer
sufficient as was first shown in \ocite{Bro}; see Fig.\ \ref{fig3} for an
example of an additional diagram analyzed in this case. The main complication
arising in \ocite{Bro} is then to treat these encounters. To simplify
the calculation we imagine these encounters being built up out of several
encounters, each of which consist of two encounter stretches.  We have distinguished
these 2-encounters by different boxes in Fig.\ \ref{fig4}.  In this way it is
much easier to consider encounter diagrams of arbitrary complexity with finite
Ehrenfest time, which did not appear in the formalism used in \ocite{Bro}. 

We first illustrate our procedure by considering three correlated orbits with
two 2-encounters as in Fig.\ \ref{fig4} and show how the result given in
\ocite{Bro} can be obtained in this case and then we treat the general case of $n$
orbits with $\left( n-1\right)$ independent or overlapping 2-encounters.

\subsection{Explanation of our procedure for $n=3$}

In the treatment of the RMT-type contribution (\ref{eq5}) we first consider the
case in which all the encounters occur inside the system.  For $n=3$ we have the
two semiclassical diagrams in Fig.\ \ref{fig1}b,c which include a trajectory set
(of three original trajectories and three partners) with two 2-encounters in
Fig.\ \ref{fig1}b and a single 3-encounter in Fig.\ \ref{fig1}c.  By shrinking
the link connecting the two encounters in Fig.\ \ref{fig1}b we can see how we
deform them into the diagram in Fig.\ \ref{fig1}c and we use this idea in our
Ehrenfest-time treatment.

\subsubsection{Two 2-encounters}

\begin{figure}
\begin{center}
\includegraphics[width=0.8\columnwidth]{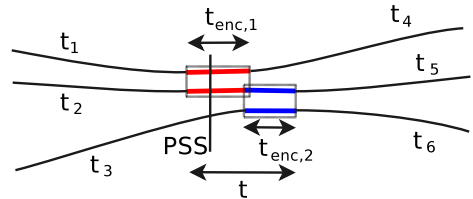}
\caption{(Color online) A diagram with two 2-encounters as we treat it with
Ehrenfest time. The encounter stretches of the two 2-encounters are marked by
boxes (shown red and blue). A possible position of the Poincar\'e surface of
section (PSS) is marked by a black vertical line.}
\label{fig4}
\end{center}
\end{figure}

For the calculation of contributions resulting from diagrams differing in
encounters we first need to review the notation and the important steps of the
corresponding calculation in \ocite{Mul1}. An encounter of two orbits is
characterized by the difference of the stable and unstable coordinates $s_i$ and
$u_i$ measured in a Poincar\'e surface of section (PSS) put inside the
encounter; see Fig.\ \ref{fig4}. In terms of these coordinates the duration of
the encounters is given by $t_{{\rm
enc},i}=1/\lambda\ln\left(c^2/\left|s_iu_i\right|\right)$ derived from the
condition that the coordinates $s_i, u_i$ are only allowed to grow up to a
classical constant $c$ (which is later related to the Ehrenfest time). The
weight function measuring the probability to find these encounters is obtained
by integrating over all possible positions where the encounter stretches can be
placed and dividing by the volume of available phase space (in the corresponding
closed system) $\Omega$ and further by the durations of the encounters $t_{{\rm
enc},i}$ to avoid overcounting the same set of correlated trajectories. The
action difference $\Delta S$ between the orbits is in general given by a
quadratic form of the coordinates $s_i, u_i$ determined by where the partner
trajectories must pierce the PSS's to reconnect in the right way to
form a closed cycle. For example for a 3-encounter one obtains \cite{Mul1}
$\Delta S=s_1'u_1'+s_2'u_2'-s_1'u_2'$ where the prime denotes that the
coordinates are measured in \textit{one} PSS from the central trajectory. If we
instead measure the coordinates in two \textit{different} sections, we obtain
$\Delta S=s_1u_1+s_2u_2-s_1u_2\exp{(-\lambda\Delta t)}$ where the time $\Delta
t$ denotes the time the particle needs to travel between the two sections. This
leads in the limit of well separated encounters to $\Delta S\approx
s_1u_1+s_2u_2$. From this and from \ocite{Mul1} we can draw the following
conclusions for the form of the action difference in the case of an arbitrary
number of (possibly overlapping) 2-encounters: In the case of $k$ well separated
2-encounters we obtain for the action difference $\Delta S\approx
\sum_{i=1}^ks_iu_i$.  When these encounters overlap the action difference can
differ from the last expression by terms exponentially damped with the time
difference between the two sections.

In our treatment, the overall contribution $C^{\ref{fig4}}(\epsilon,\tau,3)$ of
the two 2-encounters (depicted in more detail in Fig.\ \ref{fig4}) is obtained
by allowing the upper trajectory to possess a minimal length of the first
2-encounter and the lowest one a minimal length of the second 2-encounter.  The
middle trajectory, which passes through both encounters has a minimal length
given by the maximum of the two encounter times as we allow the encounters to
overlap.  However we do not yet allow one encounter to be subsumed into the
other so we also set the time $t$ between the start of the first encounter and
the end of the second to be longer than the maximum encounter time.  To write
down the semiclassical contribution of the diagram in Fig.\ \ref{fig4} we sum
over the number of possible classical orbits using the open sum rule \cite{Ric}.
Converting the time integrals resulting from this rule to time integrals with respect to
link durations, we obtain
\begin{eqnarray}
\label{eq8}
C^{\ref{fig4}}(\epsilon,\tau,3)&=& \frac{N^{2}}{\tD^{3}}\left(\prod_{i=1}^{6}
\int_{0}^{\infty}{\rm d}t_{i}{\rm e}^{-t_{i}\left(1-{\rm
i}\epsilon\right)/\tD}\right) \\
& & \times \int_{-c}^c{\rm d}^2s{\rm d}^2u\frac{{\rm e}^{{\rm
i}\epsilon\left(t_{{\rm enc},1}+t_{{\rm enc},2} \right)/\tD}}{\Omega^2t_{{\rm
enc},1}t_{{\rm enc},2}} \nonumber \\
& & \times \int_{\max \left\lbrace t_{{\rm enc},1},t_{{\rm enc},2}\right\rbrace
}^\infty \hspace{-2em} {\rm d}t\, {\rm e}^{\left( {\rm i}/\hbar\right)\Delta
S}{\rm e}^{-t\left(1-{\rm i}\epsilon\right)/\tD} , \nonumber 
\end{eqnarray}
where the superscript refers to Fig.\ \ref{fig4}. We have summed over the
possible channels, and $t_{i}$ with $i=1,\ldots,6$ label the links from the
channels to the encounters. In \eref{eq8} where ${\rm d}^2s={\rm d}s_{1}{\rm
d}s_{2}$ and ${\rm d}^2u={\rm d}u_{1}{\rm d}u_{2}$, $s_i$ and $u_i$ with $i=1,2$
are the stable and unstable coordinate differences between the two parts of the
trajectories piercing through a PSS placed in the $i$-th encounter. As explained
above, the action difference is given by $\Delta
S=s_1u_1+s_2u_2-s_1u_2\exp{(-\lambda\Delta t)}$. By expanding the part of the
exponential ${\rm e}^{({\rm i}/\hbar)\Delta S}$ containing this $\Delta
t$-dependent part  into a Taylor series one verifies easily that contributions
from higher order terms than the leading (time independent) one are of higher
order in $1/\left(\lambda\tau_D\right)$ and can be neglected. This reasoning
also holds for diagrams with more than two 2-encounters.

In the first line of \eref{eq8} we can see that each integral over the links is
weighted by its classical probability to remain inside the system for the time
$t_{i}$ which decays exponentially with the average dwell time $\tD$.  We only
want to consider trajectory sets where the \emph{whole} diagram remains inside
the system, as if any parts were to hit the lead and escape the diagram would be
truncated at that point.  With the energy dependence in \eref{eq6a} this gives
the factors ${\rm e}^{-t_{i}\left(1-{\rm i}\epsilon\right)/\tD}$ in \eref{eq8}. 
Inside the encounters however we have trajectory stretches that are so close
that the conditional survival probability of secondary traversals is 1 and we
need only consider the survival probability of one stretch.  If that stretch
does not escape then neither will the other.  The energy dependence still
depends on the total time so that encounter 1 would lead to the factor ${\rm
e}^{-t_{{\rm enc},1}\left(1-2{\rm i}\epsilon\right)/\tD}$.  With the overlap,
encounter 2 would then have a more complicated exponential factor, but because
the time $t$ (between the two outer ends of the encounter stretches on the
middle trajectory shown in Fig.\ \ref{fig4}) passes through both encounters
their survival probability (of both stretches of both encounters) can be
expressed as the survival probability of a stretch of duration $t$ as in the
last line of \eref{eq8}.  The energy dependence instead also requires the extra
traversal of the encounters as given by the exponential factor in the middle
line of \eref{eq8}.

Performing the integrals in the first line of \eref{eq8} we have
\begin{equation}
\label{eq9}
C^{\ref{fig4}}(\epsilon,\tau,3)= \frac{\tD\TH^2}{(1-{\rm i}\epsilon)^{6}}
F^{\ref{fig4}}(\tau) ,
\end{equation}
where we have moved all of the Ehrenfest-time dependent parts into the factor
$F^{\ref{fig4}}(\tau)$ with the superscript again referring to Fig.\ \ref{fig4},
\begin{eqnarray}
\label{eq10}
F^{\ref{fig4}}(\tau)&=&\int_{-c}^c{\rm d}^2s{\rm d}^2u\frac{{\rm e}^{\left( {\rm
i}/\hbar\right)\Delta S}{\rm e}^{{\rm i}\epsilon\left(t_{{\rm enc},1}+t_{{\rm
enc},2} \right)/\tD}}{\Omega^2t_{{\rm enc},1}t_{{\rm enc},2}} \nonumber \\
& & \times \int_{\max \left\lbrace t_{{\rm enc},1},t_{{\rm enc},2}\right\rbrace
}^\infty \hspace{-2em} {\rm d}t\, {\rm e}^{-t\left(1-{\rm i}\epsilon\right)/\tD}
.
\end{eqnarray}
Here we can also see the connection with the previous Ehrenfest-time treatment
of such a diagram.  When $t>t_{{\rm enc},1}+t_{{\rm enc},2}$ the two encounters
separate (the integrals can then be further broken down into products) and this
is the case in which the trajectories can be considered to have two independent
2-encounters as in \ocite{Bro}.  Because we choose a different lower limit
however, the contribution above also includes some of the diagrams previously
treated as 3-encounters in \ocite{Bro}.  The reason for our choice becomes clear
in the following steps.  We first substitute $t'=t-\max \left\lbrace t_{{\rm
enc},1},t_{{\rm enc},2}\right\rbrace$,
\begin{eqnarray}
F^{\ref{fig4}}(\tau)&=&\int_{-c}^c{\rm d}^2s{\rm d}^2u\frac{{\rm e}^{\left( {\rm
i}/\hbar\right)\Delta S}{\rm e}^{{\rm i}\epsilon\left(t_{{\rm enc},1}+t_{{\rm
enc},2} \right)/\tD}}{\Omega^2t_{{\rm enc},1}t_{{\rm enc},2}} \\
& & \times \int_{0}^\infty {\rm d}t'\, {\rm e}^{-(t'+\max \left\lbrace t_{{\rm
enc},1},t_{{\rm enc},2}\right\rbrace)\left(1-{\rm i}\epsilon\right)/\tD} ,
\nonumber
\end{eqnarray}
and then substitute $u_i=c/\sigma_i$, $s_i=cx_i\sigma_i$ and perform the
$\sigma_i$-integrals using the explicit form of the $t_{{\rm
enc},i}=1/\lambda\ln\left(c^2/\left|s_iu_i\right|\right)$ (for details of this
calculation, see also \ocite{Bro}).  This results in
\begin{eqnarray}
\label{eq12}
F^{\ref{fig4}}(\tau)&=&16\int_0^1{\rm
d}x^2\frac{\lambda^2c^4}{\Omega^2}\cos\left( \frac{c^2}{\hbar}x_1\right)
\cos\left( \frac{c^2}{\hbar}x_2\right)\nonumber\\
 &&\times \int_0^\infty {\rm d}t'\,{\rm e}^{-\left(t'+\max\left\lbrace -\ln
x_1,-\ln x_2\right\rbrace/\lambda\right)\left(1-{\rm
i}\epsilon\right)/\tD}\nonumber\\
 &&\times{\rm e}^{-{\rm i}\epsilon\left(\ln x_1+\ln x_2\right)/(\lambda\tD)} .
\end{eqnarray}
Now we substitute $x_i'=x_ic^2/\hbar$ and obtain
\begin{eqnarray}
\label{eq13}
F^{\ref{fig4}}(\tau)&=&16\int_0^{\infty}{\rm
d}{x'}^2\frac{\lambda^2\hbar^2}{\Omega^2}\cos\left(x'_1\right) \cos\left(
x'_2\right)\nonumber\\
 &&\times \int_0^\infty {\rm d}t'\,{\rm e}^{-\left(t'+\max\left\lbrace -\ln
x'_1,-\ln x'_2\right\rbrace/\lambda\right)\left(1-{\rm
i}\epsilon\right)/\tD}\nonumber\\
 &&\times{\rm e}^{-{\rm i}\epsilon\left(\ln x'_1+\ln
x'_2\right)/(\lambda\tD)}{\rm e}^{-\tau\left(1-3{\rm i}\epsilon \right)} .
\end{eqnarray}
Here we split the resulting expression into an $\hbar$-independent integral (or
more exactly trivially dependent on $\hbar$), that exists due to the energy
average that is always contained in our calculations, and an Ehrenfest-time or
$\hbar$ dependent part with $\tE\equiv1/\lambda\ln\left(c^2/\hbar \right)$. 
This contains the Ehrenfest-time dependence that is expected from \eref{eq5}, so
(\ref{eq13}) already shows that the diagrams considered here yield the correct
Ehrenfest-time dependence.

\subsubsection{A 3-encounter}

\begin{figure}
\begin{center}
\includegraphics[width=0.8\columnwidth]{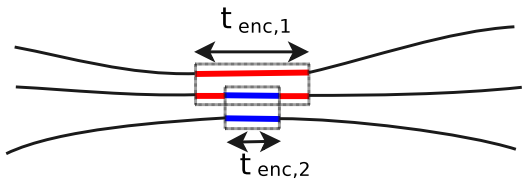}
\caption{(Color online) One 2-encounter is located fully inside the other,
corresponding to our treatment of a generalized version of a 3-encounter. The
two 2-encounters are marked by boxes (indicated by different colors). }
\label{fig5}
\end{center}
\end{figure}

Now we consider the case in which one of the two 2-encounters lies fully inside the
other one, which we will refer to as a generalized version of a 3-encounter, as
depicted in Fig.\ \ref{fig5}.

For the Ehrenfest-time dependent part we have a similar contribution as in
\eref{eq10} with two differences:  First $t$ is best defined as the distance
between the midpoints of the two different encounter stretches and so it can vary
between 
\begin{eqnarray}
\vert t \vert &\leq& \frac{1}{2}\left(\max\left\lbrace t_{{\rm enc},1},t_{{\rm
enc},2}\right\rbrace-\min\left\lbrace t_{{\rm enc},1},t_{{\rm
enc},2}\right\rbrace\right) , \nonumber \\
\vert t \vert &\leq& \frac{1}{2}\vert t_{{\rm enc},1} - t_{{\rm enc},2}\vert .
\end{eqnarray}
Second the survival probability of the encounters is determined by the longest
encounter stretch and is independent of $t$.  The Ehrenfest-time dependent part
can then be written as
\begin{eqnarray}
F^{\ref{fig5}}(\tau)&=&\int_{-c}^c{\rm d}^2s{\rm d}^2u\frac{{\rm e}^{\left( {\rm
i}/\hbar\right)\Delta S}{\rm e}^{{\rm i}\epsilon\left(t_{{\rm enc},1}+t_{{\rm
enc},2} \right)/\tD}}{\Omega^2t_{{\rm enc},1}t_{{\rm enc},2}}  \\
& & \times \int_{-\frac{1}{2}\vert t_{{\rm enc},1} - t_{{\rm
enc},2}\vert}^{\frac{1}{2}\vert t_{{\rm enc},1} - t_{{\rm enc},2}\vert}
\hspace{-2em}{\rm d}t\, {\rm e}^{-(\max \left\lbrace t_{{\rm enc},1},t_{{\rm
enc},2}\right\rbrace)\left(1-{\rm i}\epsilon\right)/\tD} . \nonumber
\end{eqnarray}
Performing the $t$ integral and following the same steps as for
\erefs{eq12}{eq13}, we find
\begin{eqnarray}
\label{eq16}
F^{\ref{fig5}}(\tau)&=&16\int_0^{\infty}{\rm
d}{x'}^2\frac{\lambda^2\hbar^2}{\Omega^2}\frac{\vert \ln x'_1 - \ln x'_2
\vert}{\lambda}\cos\left(x'_1\right) \nonumber\\
 &&\times \cos\left( x'_2\right){\rm e}^{-\left(\max\left\lbrace -\ln x'_1,-\ln
x'_2\right\rbrace\right)\left(1-{\rm i}\epsilon\right)/(\lambda\tD)}\nonumber\\
 &&\times{\rm e}^{-{\rm i}\epsilon\left(\ln x'_1+\ln
x'_2\right)/(\lambda\tD)}{\rm e}^{-\tau\left(1-3{\rm i}\epsilon \right)} .
\end{eqnarray}
This part also shows an Ehrenfest-time dependence as expected from \eref{eq5}.
Note that when performing the $t$-integral the result in this case is of course
proportional to $\vert t_{{\rm enc},1} - t_{{\rm enc},2}\vert$ which contains,
after the substitution from $x$ to $x'$, two times the same terms linear in
$\tE$ with different signs that thus cancel each other.

\subsubsection{Touching the lead}

Up to now we have concentrated on encounters inside the system, but apart from these
diagrams we also need to consider diagrams where the encounters touch the
opening as in Fig.\ \ref{fig1}d-h. We will, as above, start by considering
encounters built up out of two 2-encounters and we focus here on how the
calculation of the contribution is changed when encounters move into the lead
compared to the treatment of encounters inside the system.  As can also be found
in more detail in \ocite{Bro} when encounters touch the lead one includes in the
semiclassical expressions for encounters inside the system an additional time
integral running between zero and the corresponding encounter time, which
characterizes the duration of the part of the encounter stretch that has not yet
been moved into the lead.

\begin{figure}
\begin{center}
\includegraphics[width=0.5\columnwidth]{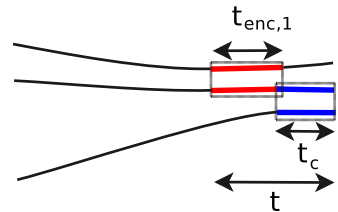}
\caption{(Color online) The second of two 2-encounters now enters the lead so
that only $t_c$ of it remains inside the system.}
\label{fig6}
\end{center}
\end{figure}

We consider two encounters with durations $t_{{\rm enc},1}$ and $t_{{\rm
enc},2}$ with the second encounter touching the opening as in Fig.\ \ref{fig1}d
and drawn in more detail in Fig.\ \ref{fig6}.  As the second encounter enters
the lead we now define the time $t$ to be from the start of the first encounter
until the lead and introduce the time $t_c$ which measures the part of the
second encounter that has not yet been moved into the lead.  We also separate
the Ehrenfest-time relevant contribution $F^{\ref{fig6}}(\tau)$ in this detailed
calculation into two cases: in the first case (A); $t_{{\rm enc},2}<t_{{\rm
enc},1}$; we have $F_{{\rm A}}^{\ref{fig6}}(\tau)$ with the additional integral
over the time $t_c$
\begin{eqnarray}
\label{eq17}
F_{{\rm A}}^{\ref{fig6}}(\tau)&=&\int_{\begin{subarray}{l}-c \\ t_{{\rm
enc},2}<t_{{\rm enc},1}\end{subarray}}^c\hspace{-2em}{\rm d}^2s{\rm
d}^2u\frac{{\rm e}^{\left({\rm i}/\hbar\right)\Delta S}{\rm e}^{{\rm i}\epsilon
t_{{\rm enc},1}/\tD}}{\Omega^2t_{{\rm enc},1}t_{{\rm enc},2}} \\
&&\times \int_0^{t_{{\rm enc},2}}{\rm d}t_c\, {\rm e}^{{\rm i}\epsilon
t_{c}/\tD}\int_{t_{{\rm enc},1}}^\infty {\rm d}t\, {\rm e}^{-t(1-{\rm
i}\epsilon)/\tD} , \nonumber
\end{eqnarray}
where the limits on the time integrals derive from the fact that the first
encounter is not allowed to touch the lead (this would be included as a
3-encounter) and that the second must.  Performing the time integrals this is
\begin{eqnarray}
F_{{\rm A}}^{\ref{fig6}}(\tau)&=&\int_{\begin{subarray}{l}-c \\ t_{{\rm
enc},2}<t_{{\rm enc},1}\end{subarray}}^c\hspace{-2em}{\rm d}^2s{\rm
d}^2u\frac{{\rm e}^{\left({\rm i}/\hbar\right)\Delta S}}{\Omega^2t_{{\rm
enc},1}t_{{\rm enc},2}}\frac{\tD^2}{{\rm i}\epsilon(1-{\rm
i}\epsilon)}\nonumber\\
&&\times \left[{\rm e}^{{\rm i}\epsilon t_{{\rm enc},2}/\tD}-1\right]{\rm
e}^{-t_{{\rm enc},1}(1-2{\rm i}\epsilon)/\tD} ,
\end{eqnarray}
with the first and second term in the square brackets resulting from the upper
and lower limit of the $t_c$-integration. In the second case (B); $t_{{\rm
enc},2}>t_{{\rm enc},1}$; we obtain
\begin{eqnarray}
\label{eq19}
F_{{\rm B}}^{\ref{fig6}}(\tau)&=&\int_{\begin{subarray}{l}-c \\ t_{{\rm
enc},2}>t_{{\rm enc},1}\end{subarray}}^c\hspace{-2em}{\rm d}^2s{\rm
d}^2u\frac{{\rm e}^{\left({\rm i}/\hbar\right)\Delta S}{\rm e}^{{\rm i}\epsilon
t_{{\rm enc},1}/\tD}}{\Omega^2t_{{\rm enc},1}t_{{\rm enc},2}} \\
&&\times \left[ \int_0^{t_{{\rm enc},1}}{\rm d}t_c\, {\rm e}^{{\rm i}\epsilon
t_{c}/\tD}\int_{t_{{\rm enc},1}}^\infty {\rm d}t\, {\rm e}^{-t(1-{\rm
i}\epsilon)/\tD} \right .\nonumber \\
&& {} + \left. \int_{t_{{\rm enc},1}}^{t_{{\rm enc},2}}{\rm d}t_c\, {\rm
e}^{{\rm i}\epsilon t_{c}/\tD}\int_{t_{c}}^\infty {\rm d}t\, {\rm e}^{-t(1-{\rm
i}\epsilon)/\tD} \right] , \nonumber
\end{eqnarray}
where the more complicated limits derive from not allowing the second encounter
to move further left than the first.  After integrating we have
\begin{eqnarray}
F_{{\rm B}}^{\ref{fig6}}(\tau)&=&\int_{\begin{subarray}{l}-c \\ t_{{\rm
enc},2}>t_{{\rm enc},1}\end{subarray}}^c\hspace{-2em}{\rm d}^2s{\rm
d}^2u\frac{{\rm e}^{\left({\rm i}/\hbar\right)\Delta S}}{\Omega^2t_{{\rm
enc},1}t_{{\rm enc},2}}\frac{\tD^2}{(1-{\rm i}\epsilon)}\\
&&\times \left[ \frac{1}{{\rm i}\epsilon} \left[{\rm e}^{{\rm i}\epsilon t_{{\rm
enc},1}/\tD}-1\right]{\rm e}^{-t_{{\rm enc},1}(1-2{\rm i}\epsilon)/\tD} \right .
\nonumber\\
&& {} + \frac{1}{(1-2{\rm i}\epsilon)}{\rm e}^{-t_{{\rm enc},1}(1-3{\rm
i}\epsilon)/\tD} \nonumber \\
& & {} - \left. \frac{1}{(1-2{\rm i}\epsilon)}{\rm e}^{{\rm i}\epsilon t_{{\rm
enc},1}/\tD}{\rm e}^{-t_{{\rm enc},2}(1-2{\rm i}\epsilon)/\tD} \right ].
\nonumber
\end{eqnarray}
The last line comes from the upper limit of the second $t_c$-integral and has
the same Ehrenfest-time dependence as before and in line with \eref{eq5}. 
Likewise the upper $t_c$ time limit for case A in \eref{eq17} leads to the same
dependence and we can conclude that the upper limits of the $t_c$-integrations
yield contributions similar to when the encounters are inside the system and
with the same Ehrenfest-time dependence. The remaining (lower) limits of the
time integrations in \erefs{eq17}{eq19} give contributions possessing a
different Ehrenfest-time dependence which however always yield zero in the
semiclassical limit due to the fact that the corresponding terms contain no
$t_{{\rm enc},2}$ in the exponentials containing $\tD$.   Apart from the action
difference, the only term depending on $s_2,u_2$ is the $1/t_{{\rm enc},2}$. The
resulting expression is rapidly oscillating as a function of the energy
\cite{Mul1} and thus canceled by the energy average.

We can repeat this procedure for the remaining diagrams in Fig.\ \ref{fig1} and
see that the contributions are determined by the upper limits of the
corresponding $t_c$ integrals.  For the diagrams with a generalized 3-encounter
(Fig.\ \ref{fig1}g,h) this follows as for the 3-encounter inside the system
but for Fig.\ \ref{fig1}e where the two 2-encounters enter different channels
(and possibly different leads) there is an additional subtlety.  The two
encounters are still allowed to overlap, so that during the time $t$ the stretch
now connecting both channels can always be inside encounters but the individual
encounters are not allowed to connect leads at both ends.  These additional
possibilities are considered later, where if both encounters connect to the
leads at both ends we actually have a band of correlated trajectories (treated
in Sec.\ \ref{part2}) and if only one does we have a mixed term (treated in
Sec.\ \ref{mixing}).  With this organization of the encounters we see that
each diagram has the same Ehrenfest-time dependence as when the encounters are
inside the system, which is in line with \eref{eq5}.

\subsubsection{Intermediate summary}

The reasoning so far in this section proves the form of \eref{eq5} for $n=3$.
First of all, we know that the resulting contribution from the diagrams analyzed
contains an overall factor ${\rm e}^{-\tau(1-3{\rm i}\epsilon)}$.  Secondly, the
remaining integrals are independent of $\hbar$ and thus independent of the
Ehrenfest time.  Thirdly, the diagrams we analyze are the same as the ones
analyzed in the RMT-case in the first part of \ocite{Kui}.  As in the limit
$\tE\to0$ we must recover that previous result, this implies that
$C(\epsilon,\tau,3)$ in \eref{eq5} is indeed given by the RMT-expression.

\subsubsection{Full contributions}

Before proceeding to the general case however, we first want to illustrate how
our calculation can be used to obtain, apart from just the Ehrenfest-time
dependence, the complete dependence on $\tD$ and $\epsilon$.

We therefore start for the two 2-encounters from Fig.\ \ref{fig4} from the last
expression in \eref{eq13} and perform first the $t'$-integral
\begin{eqnarray}
F^{\ref{fig4}}(\tau)&=&\frac{16\tD}{\left(1-{\rm
i}\epsilon\right)}\int_0^{\infty}{\rm
d}{x'}^2\frac{\lambda^2\hbar^2}{\Omega^2}\cos\left(x'_1\right) \cos\left(
x'_2\right)\nonumber\\
 &&\times {\rm e}^{-\max\left\lbrace -\ln x'_1,-\ln
x'_2\right\rbrace\left(1-2{\rm i}\epsilon\right)/(\lambda\tD)}\nonumber\\
 &&\times{\rm e}^{\min\left\lbrace -\ln x'_1,-\ln x'_2\right\rbrace {\rm
i}\epsilon/(\lambda\tD)}{\rm e}^{-\tau\left(1-3{\rm i}\epsilon \right)} ,
\end{eqnarray}
where it is simpler to rewrite the result in terms of the maximum and minimum
value of $\ln x'_{i}$.  For calculating the $x'_i$-integrals we perform partial
integrations (integrating each time the $\cos$ functions) and then perform the
resulting integrals from zero to infinity
\begin{eqnarray}
\label{eq22}
F^{\ref{fig4}}(\tau)&=&-\frac{16{\rm i}\epsilon}{\tD}\frac{\left(1-2{\rm
i}\epsilon\right)}{\left(1-{\rm i}\epsilon\right)}\int_0^{\infty}{\rm
d}{x'}^2\frac{\hbar^2}{\Omega^2}\frac{\sin\left(x'_1\right)}{x'_{1}}
\frac{\sin\left( x'_2\right)}{x'_{2}}\nonumber\\
 &&\times {\rm e}^{-\max\left\lbrace -\ln x'_1,-\ln
x'_2\right\rbrace\left(1-2{\rm i}\epsilon\right)/(\lambda\tD)}\nonumber\\
 &&\times{\rm e}^{\min\left\lbrace -\ln x'_1,-\ln x'_2\right\rbrace {\rm
i}\epsilon/(\lambda\tD)}{\rm e}^{-\tau\left(1-3{\rm i}\epsilon \right)}
\nonumber \\
& = & -\frac{{\rm i}\epsilon}{\tD\TH^{2}}\frac{\left(1-2{\rm
i}\epsilon\right)}{\left(1-{\rm i}\epsilon\right)}{\rm e}^{-\tau\left(1-3{\rm
i}\epsilon \right)} .
\end{eqnarray}
In the first line the additional terms due the partial integration are either zero
or cancel due to the energy average.  The final result in the last line of
\eref{eq22} can be also obtained by replacing $\max\left\lbrace -\ln x_1',-\ln
x_2'\right\rbrace/\lambda = y_1$ and $\min\left\lbrace -\ln x_1',-\ln
x_2'\right\rbrace/\lambda = y_2$ and performing the integrals with respect to
$y_i$ from zero to infinity.

To evaluate the contribution from the generalized 3-encounter in Fig.\
\ref{fig5} we again perform two partial integrations in \eref{eq16} and obtain
\begin{eqnarray}
\label{eq23}
F^{\ref{fig5}}(\tau)&=&\frac{16}{\tD}\left(1-{\rm
i}\epsilon\right)\int_0^{\infty}{\rm
d}{x'}^2\frac{\hbar^2}{\Omega^2}\frac{\sin\left(x'_1\right)}{x'_{1}}
\frac{\sin\left( x'_2\right)}{x'_{2}}\nonumber\\
 &&\times {\rm e}^{-\max\left\lbrace -\ln x'_1,-\ln
x'_2\right\rbrace\left(1-2{\rm i}\epsilon\right)/(\lambda\tD)}\nonumber\\
 &&\times{\rm e}^{\min\left\lbrace -\ln x'_1,-\ln x'_2\right\rbrace {\rm
i}\epsilon/(\lambda\tD)}{\rm e}^{-\tau\left(1-3{\rm i}\epsilon \right)}
\nonumber \\
& = & \frac{\left(1-{\rm i}\epsilon\right)}{\tD\TH^{2}}{\rm
e}^{-\tau\left(1-3{\rm i}\epsilon \right)} ,
\end{eqnarray}
where we have also left out the terms from the partial integrations that cancel
due to the energy average.

With these results we can now show how they connect to the RMT-type results. 
For this we need to split our diagrams differently and first we need the result for
an ideal 3-encounter as depicted in Fig.\ \ref{fig2} whose contribution was
calculated \cite{Bro} to be
\begin{equation}
\label{eq24}
F^{\ref{fig2}}(\tau)=-\frac{\left(1-3{\rm i}\epsilon\right)}{\tD\TH^{2}}{\rm
e}^{-\tau\left(1-3{\rm i}\epsilon \right)} .
\end{equation}
With the extra factors in \eref{eq9} it is clear how in the limit $\tE=0$ this
reduces to the RMT-type result for a 3-encounter as in \ocite{Kui}.  All the
remaining contributions should be collected together as two 2-encounters, and as
the ideal 3-encounter is included in our generalized 3-encounter we first
subtract \eref{eq24} from \eref{eq23}
\begin{equation}
\label{eq25}
F^{\ref{fig5}}(\tau)-F^{\ref{fig2}}(\tau)=2\frac{\left(1-2{\rm
i}\epsilon\right)}{\tD\TH^{2}}{\rm e}^{-\tau\left(1-3{\rm i}\epsilon \right)}.
\end{equation}
Before we add the result from our separation of two 2-encounters in \eref{eq22}
we remember that in the treatment we enforce that the first encounter is to the
left of the second.  The result in \eref{eq25} does not have this restriction so
we divide by 2 to ensure compatibility and then add the result in \eref{eq22} to
obtain
\begin{equation}
\label{eq26}
F^{\ref{fig3}}(\tau)=\frac{1}{\tD\TH^{2}}\frac{\left(1-2{\rm
i}\epsilon\right)^{2}}{\left(1-{\rm i}\epsilon\right)}{\rm
e}^{-\tau\left(1-3{\rm i}\epsilon \right)} .
\end{equation}
This then reduces to the RMT-type result for trajectories with two 2-encounters
when $\tE=0$ as in \ocite{Kui}.  The agreement of these results with the
previous Ehrenfest-time treatment\cite{Bro} can be seen as the result in
\eref{eq26} including both the result from two independent 2-encounters as well
as most of the contribution of the diagram referred to as a 3-encounter in
\ocite{Bro}. When splitting the contribution in a different way as in
\ocite{Bro} this also leads to terms in both classes that contain different
Ehrenfest-time dependencies that only cancel when summed together.

\subsection{All orders}

Although up to now we have just reproduced results from \ocite{Bro}, the
procedure used here has the advantage that it yields a simple algorithm for
determining the Ehrenfest-time dependence of the corresponding contributions to
$C_{1}(\epsilon,\tau,n)$ at arbitrary order.  For our example of $n=3$ we showed
how it was possible to split the diagrams into two classes that \textit{both}
showed the Ehrenfest-time dependence as expected from \eref{eq5}.  We want now to
show how to generalize our way of splitting considered for three trajectories
to diagrams containing $n$ trajectories.

\subsubsection{Ladder diagrams}

\begin{figure}
\begin{center}
\includegraphics[width=0.8\columnwidth]{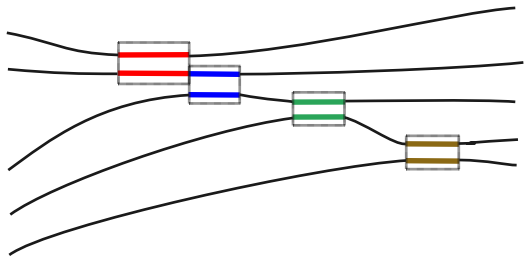}
\caption{(Color online) A ladder of consecutive 2-encounters. The encounter
stretches are marked by boxes (shown in different colors).}
\label{fig7}
\end{center}
\end{figure}

We start again with the situation in which all of the encounters are inside the
system and by considering a case analogous to Fig.\ \ref{fig4}, but now involving
$n$ instead of three trajectories. We first take a diagram that consists of
a ladder of $(n-1)$ 2-encounters so that the central $n-2$ trajectories each
contain two encounter stretches while the two outside trajectories only contain
one encounter stretch each.  This situation is depicted in Fig.\ \ref{fig7} and
the encounters are thus characterized by $(n-1)$ $s,u$-coordinates.

In this case we obtain for the Ehrenfest-time relevant contribution
$F^{\ref{fig7}}(\tau)$ that the $t$-integral measuring the time difference
between the end points of the two encounter stretches on the middle orbit in
\eref{eq10} is replaced by $n-2$ integrals over times $t_i$ with the same
meaning as $t$; they measure the time difference between the end points of the
two (consecutive) encounter stretches on the central trajectories containing two
encounter stretches.  These times likewise run from the maximum of the
corresponding encounter times to infinity.  The survival probability is
determined by a single (artificial) stretch that runs through all the encounters
so that the exponential term describing the $\tD$- and $\epsilon$-dependence is
now given by
\begin{equation}
\label{ladderexponentialeqn}
{\rm e}^{-\sum_{i=1}^{n-2}t_i \left(1-{\rm i}\epsilon\right)/\tau_D}{\rm
e}^{\sum_{i=2}^{n-2}t_{{\rm enc},i}/\tD}{\rm e}^{{\rm i}\epsilon\left(t_{{\rm
enc},1}+t_{{\rm enc},n-1}\right)/\tD } ,
\end{equation}
where $t_{{\rm enc},i}$ are the durations of the $(n-1)$ individual 2-encounters
and the middle exponential compensates for the fact that the middle encounters
are traversed by two $t_{i}$ and that only one traversal should contribute to
the survival probability.  Setting $t'_{i}=t_{i}-\max \left\lbrace t_{{\rm
enc},i},t_{{\rm enc},i+1}\right\rbrace$ and repeating now the steps of
\erefs{eq12}{eq13} we find the Ehrenfest-time dependent factor in this case to
be
\begin{eqnarray}
\label{lad}
F^{\ref{fig7}}(\tau)&=&\left(\frac{4\lambda\hbar}{\Omega}\right)^{n-1}\prod_{j=1
}^{n-1}\int_0^{\infty}{\rm
d}{x'_{j}}\cos\left(x'_j\right)\prod_{i=1}^{n-2}\int_0^\infty {\rm
d}t'_{i}\nonumber\\
 &&\times {\rm e}^{-\sum_{i=1}^{n-2}\left(t'_{i}+\max\left\lbrace -\ln
x'_{i},-\ln x'_{i+1}\right\rbrace/\lambda\right)\left(1-{\rm
i}\epsilon\right)/\tD}\nonumber\\
 &&\times{\rm e}^{-\sum_{i=2}^{n-2}\ln x'_{i}/(\lambda\tD)}{\rm e}^{-{\rm
i}\epsilon\left(\ln x'_1+\ln x'_{n-1}\right)/(\lambda\tD)}\nonumber \\
 && \times{\rm e}^{-\tau\left(1-{\rm i} n\epsilon \right)} ,
\end{eqnarray}
again confirming the Ehrenfest-time dependence of \eref{eq5}.

\subsubsection{Single encounter}

Along with the case in which none of the encounters in the ladder can move
completely inside another, we can look at the opposite extreme where all the
encounter stretches lie inside of the encounter $k$ with the longest duration
$t_{{\rm enc},k}=\max_i\left\lbrace t_{{\rm enc},i}\right\rbrace$ where $t_{{\rm
enc},i}$ are the durations of the $(n-1)$ individual 2-encounters with one of
the two orbits containing the stretch of duration $t_{{\rm enc},k}$.  This
situation is like a generalization of the diagram in Fig.\ \ref{fig5} and we
similarly now define the times $t_{i}$ to be measured between the centers of
encounter $i$ and the encounter $k$ of maximum length (with $i\neq k$).  Here
the same Ehrenfest-time dependence ${\rm e}^{-\tau\left(1-{\rm i} n\epsilon
\right)}$ follows by taking into account that each time $t_i$ has a range of
variation of size $t_{{\rm enc},k}-t_{{\rm enc},i}$ and that the $\tD$- and
$\epsilon$-dependent exponential in this case is
\begin{equation}
\label{singleencounterexponentialeqn}
{\rm e}^{-t_{{\rm enc},k}\left(1-{\rm i}\epsilon\right)/\tD}{\rm e}^{{\rm
i}\epsilon\sum_{i=1}^{n-1}t_{{\rm enc},i}/\tD} .
\end{equation}
This yields for the Ehrenfest-time dependent factor
\begin{eqnarray}
F^{\ref{fig7}'}(\tau)&=&\left(\frac{4\lambda\hbar}{\Omega}\right)^{n-1}\prod_{
j=1}^{n-1}\int_0^{\infty}{\rm d}{x'_{j}}\cos\left(x'_j\right)\nonumber\\
&& \times {\rm e}^{\left(1-{\rm i}\epsilon\right)\ln
x'_k/(\lambda\tD)}\left[\prod_{\substack{i=1\\i\neq k}}^{n-1}\frac{ \left(\ln
x'_i - \ln x'_k\right)}{\lambda}\right]\nonumber\\
 &&\times {\rm e}^{-{\rm i}\epsilon\sum_{i=1}^{n-1}\ln x'_{i}/(\lambda\tD)}{\rm
e}^{-\tau\left(1-{\rm i} n\epsilon \right)} ,
 \end{eqnarray}
confirming again the Ehrenfest-time dependence predicted by \eref{eq5}.

\subsubsection{Mixture}

\begin{figure}
\begin{center}
\includegraphics[width=0.8\columnwidth]{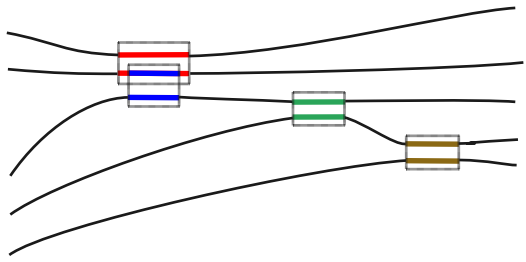}
\caption{(Color online) One possible example of a mixed case: One encounter is
fully contained inside an other, the others form a ladder as considered before.}
\label{fig7a}
\end{center}
\end{figure}

Of course it is additionally possible to have a mixed form between these two
extreme cases. This means that some 2-encounters only overlap like in the case
of a ladder diagram while the others form `single' encounters; see Fig.\
\ref{fig7a} for a possible diagram.  We then have a ladder of `combined'
encounters that themselves can be made up of one or more 2-encounters.  The
treatment of such diagrams is very similar to the treatments above, and the only
slight complication is in defining the appropriate times to extract the
Ehrenfest-time dependence.

We recall that the first and last trajectories only pass through one 2-encounter
while the central $n-2$ trajectories pass through two.  Numbering the central
trajectories from $1,\ldots,n-2$, so that trajectory $i$ has encounters $i$ and
$i+1$ along it, we divide them into two sets: those whose encounter stretches
lie fully inside each other or a connected encounter, as in the case of a single
encounter above, that we place in the set $\Omega_1$. We place the remaining orbits with two
stretches separated as in ladder diagrams in the set $\Omega_2$.  As
mentioned above, we condense the overlapping encounters into combined encounters
and record in the set $\Lambda$ the labels of the trajectories that pass
through the second stretch in each combined encounter. We also include in this
set combined encounters made of a single separated (ladder) 2-encounter. We then
use $m(i)$ for $i\in\Lambda$ to record the number of additional consecutive
trajectories involved in the same encounter, so that $m(i)=0$ for separated
2-encounters and $m(i)>0$ for larger encounters corresponding to the single
encounter case above.  If the last combined encounter is a 2-encounter its
second stretch is traversed by the last trajectory in the diagram which we
number by $n-1$ and include as an element of $\Lambda$.  For example, for the
diagram in Fig.\ \ref{fig7a} we would have $\Omega_1=\{1\}$, $\Omega_2=\{2,3\}$,
$\Lambda=\{1,3,4\}$, $m(1)=1$, $m(3)=0$ and $m(4)=0$.  For the elements
$i\in\Lambda$ we also label by $k_i$ the corresponding encounter of maximum
length among those from encounter $i$ to encounter $i+m(i)$.  To be precise, the
two stretches that stay together longest have length $t_{\mathrm{enc},k_i}$
while the other encounter times are defined by how long the remaining stretches
remain close to one of the two longest.

For the trajectories passing through two separated condensed encounters
$i\in\Omega_2$ we define the times $t_i$ to include the whole of the leftmost
and rightmost condensed encounters, i.e.\ to include the encounters
$k_{\bar{i}}$ and $k_{i+1}$, where $\bar{i}$ is the largest element in $\Lambda$
that is $\leq i$.  In this case the $\tD$- and $\epsilon$-dependent exponential
can be written as
\begin{eqnarray}
\label{mix}
&&{\rm e}^{-\sum_{i\in\Omega_2}t_i\left(1-{\rm i}\epsilon\right)/\tD}{\rm
e}^{\sum_{i\in\bar{\Lambda}} t_{{\rm enc},k_i}\left(1-{\rm
i}\epsilon\right)/\tD}\nonumber\\
&&\times{\rm e}^{-t_{{\rm enc},k_1}\left(1-{\rm i}\epsilon\right)/\tD}
{\rm e}^{{\rm i}\epsilon\sum_{i=1}^{n-1}t_{{\rm enc},i}/\tD} ,
\end{eqnarray}
where $\bar{\Lambda}$ is $\Lambda$ with its largest element removed so that the
second term accounts for the overlap between the $t_i$'s and the third term for
the fact that there is also no overlap at the start of the first such stretch. 
This equation incorporates both \eref{ladderexponentialeqn} and
\eref{singleencounterexponentialeqn}.  If we introduce the notation
$M[i,j]\equiv\max\left\lbrace t_{{\rm enc},i},t_{{\rm enc},j}\right\rbrace$, we
can then define the times $t'_{i}=t_i-M[k_{\bar{i}},k_{i+1}]$ as before
\eref{lad}. Making the substitutions as done previously yields for the Ehrenfest-time dependent factor 
\begin{eqnarray}
\label{mix1}
F^{\ref{fig7a}}(\tau)&=&\left(\frac{4\lambda\hbar}{\Omega}\right)^{n-1}\prod_{
j=1}^{n-1}\int_0^{\infty}{\rm d}{x'_{j}}\cos\left(x'_j\right)
\prod_{i\in\Omega_2}\int_0^\infty {\rm d}t_i' \nonumber\\
&&\times\prod_{i\in(\Omega_1\cap\Lambda)}\left[\prod_{\substack{j=i\\j\neq
k_i}}^{i+m(i)}\frac{ \left(\ln x'_j - \ln
x'_{k_i}\right)}{\lambda}\right]\nonumber\\
&&{\rm
e}^{-\sum_{i\in\Omega_2}\left(t'_i+\hat{M}[k_{\bar{i}},k_{i+1}]
/\lambda\right)\left(1-{\rm i}\epsilon\right)/\tD}\nonumber\\
&&\times{\rm e}^{-\sum_{i\in\bar{\Lambda}} \ln x'_{k[i]}\left(1-{\rm
i}\epsilon\right)/(\lambda\tD)}{\rm e}^{\ln x'_{k[1]}\left(1-{\rm
i}\epsilon\right)/(\lambda\tD)}\nonumber\\
&& \times{\rm e}^{-{\rm i}\epsilon\sum_{i=1}^{n-1}\ln x'_{i}/(\lambda\tD)}
{\rm e}^{-\tau\left(1-{\rm i} n\epsilon \right)} ,
\end{eqnarray}
with $\hat{M}[i,j]=\max\left\lbrace -\ln x'_{i},-\ln x'_{j}\right\rbrace$. As
$\Omega_2$ and $\bar{\Lambda}$ must have the same number of elements, this again
shows the predicted Ehrenfest-time dependence. 

\subsubsection{General encounters}

\begin{figure}
\begin{center}
\includegraphics[width=0.8\columnwidth]{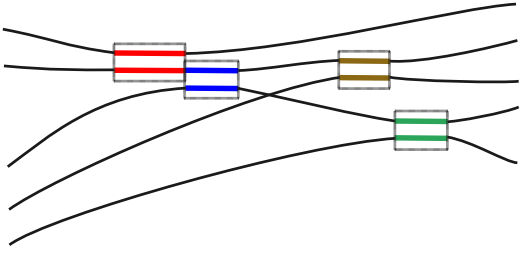}
\caption{(Color online) A general diagram containing also orbits with more than
two stretches. The encounters are marked by boxes (shown in different colors).}
\label{fig8}
\end{center}
\end{figure}

Up to now we restricted our discussion to diagrams in which each trajectory is
involved in one or two encounters. This is however not yet the most general case
where the only restriction is that each trajectory contains at least one
encounter stretch, so that some trajectories can also contain more than two
encounter stretches. Note that the situation where two trajectories interact
(pass through the same 2-encounter block) more than once cannot occur at leading
order in inverse channel number.  An example of a diagram that is possible is
depicted in Fig.\ \ref{fig8}. In the most general case we define the times $t_i$
slightly differently: first we separate the $k\geq2$ trajectories that have one
encounter stretch from the remaining $n-k$ that have more than one.  Then we
number our encounters accordingly, first those along the trajectories with one
encounter stretch with duration $t_{{\rm enc},i}$, $i=1,\ldots,k$ then the
remaining encounters with duration $t_{{\rm enc},i}$, $i=k+1,\ldots,n-1$.  For
the $n-k$ trajectories with two or more encounter stretches we now define
$t_{i}$, $i=1,\ldots,n-k$, to be the time difference between the outer edges of
the outermost encounters along those trajectories.  

For any trajectories with more than two encounter stretches we will need
additional time differences to fully fix the positions of the encounters. 
Because we defined the times $t_{i}$ to go through the outmost encounters,
importantly the exponential factor with the survival probability and the energy
dependence does not depend on these additional time differences and is given by
\begin{equation} 
{\rm e}^{-\sum_{i=1}^{n-k}t_i\left(1-{\rm i}\epsilon\right)/\tD}{\rm
e}^{\sum_{i=k+1}^{n-1}t_{{\rm enc},i}/\tD}{\rm e}^{{\rm i}\epsilon\sum_{i=1}^k
t_{{\rm enc},i}/\tD}
\end{equation}
where the middle term ensures that the survival probability only includes one
copy of each encounter and the energy dependence involves all traversals of all
the encounters.

\begin{figure}
\begin{center}
\includegraphics[width=0.65\columnwidth]{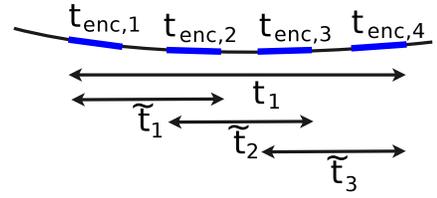}
\caption{(Color online) Definition of the times $\tilde{t}_i$ in the case of
more than two encounter stretches on one orbit. The encounter stretches are
shown thicker (blue).}
\label{fig9}
\end{center}
\end{figure}

For the remaining times we notice that, starting with the ladder system with 2
trajectories containing one encounter stretch and $n-2$ trajectories containing
two stretches, every time we increase the number of trajectories with one
encounter stretch we simultaneously increase the number with more than two. 
Therefore there are $k-2$ additional time differences needed to fix the
positions of the central encounters along trajectories with more than two and we
define times $\tilde{t}_i$ for $i=1,\ldots,k-2$ from the left hand side of one
encounter stretch to the right hand side of the next encounter stretch following
on the right on those trajectories, see also Fig.\ \ref{fig9}.  As the
encounters are ordered, they are not (yet) allowed to be subsumed by each other
or pushed past the outside encounters. The ranges of the times $\tilde{t}_i$ are
then fixed by these restrictions.  Using again $M[i,j]$ defined after \eref{mix}
in the following to make the notation more compact, we obtain for a trajectory
containing $m$ encounter stretches of durations $t_{{\rm enc},i}$, $i=1,\ldots,m$, 
as illustrated in Fig.\ \ref{fig9}, the integrals
\begin{eqnarray}
&&\int_{M(1,2)}^{t_i}{\rm
d}\tilde{t}_1\ldots\int_{M[m-2,m-1]}^{t_i-\sum_{o=1}^{m-3}\left(\tilde{t}_o-M[o,
o+1]\right)}{\rm d}\tilde{t}_{m-2}\nonumber\\ 
&=&\int_0^{t_i-M[1,2]}{\rm
d}\tilde{t}_1'\ldots\int_0^{t_i-\sum_{o=1}^{m-3}\tilde{t}_o'-M[m-2,m-1]}{\rm
d}\tilde{t}_{m-2}'.\nonumber\\
\label{eq32}
\end{eqnarray}
In the second line we substituted $\tilde{t}_j'=\tilde{t}_j-M[j,j+1]$. The time
differences $t_{i}$, which are more important for the Ehrenfest-time dependence,
must instead just be longer than the maximal length of the encounter stretches
lying on the considered trajectory.  In general the numbering of the encounters
and time differences can be more complicated than in Fig.\ \ref{fig9} so we
define $l(i)$ to be a list of length $m(i)$ of the encounters enclosed by the
time $t_{i}$ (including the outer encounters) and $L(i)$ a list of the
corresponding $m(i)-1$ times $\tilde{t}$ between the ends of those encounters. 
Now we can make the substitution $t'_{i}=t_{i}-\max_{j\in l(i)}\left\lbrace
t_{{\rm enc},j}\right\rbrace$. After this substitution we recognize that
\eref{eq32} has become independent of $\hbar$ or the Ehrenfest time. Following
then the steps in \erefs{eq12}{eq13} we obtain
\begin{eqnarray}
F^{\ref{fig8}}(\tau)&=&\left(\frac{4\lambda\hbar}{\Omega}\right)^{n-1}\prod_{j=1
}^{n-1}\int_0^{\infty}{\rm
d}{x'_{j}}\cos\left(x'_j\right)\prod_{i=1}^{n-k}\int_0^\infty {\rm
d}t'_{i}\nonumber\\
&& \times \int_0^{t_i'-\left(\ln
x'_{\max,i}+\hat{M}[l_{1},l_{2}]\right)/\lambda}{\rm
d}\tilde{t}_{L_{1}}'\ldots\nonumber\\ &&\times
\int_0^{t_i'-\sum_{o=1}^{m-3}\tilde{t}_{L_{o}}'-\left(\ln
x'_{\max,i}+\hat{M}[l_{m-2},l_{m-1}]\right)/\lambda}\hspace{-0.5em}{\rm
d}\tilde{t}_{L_{m-2}}' \nonumber \\
&& \times {\rm e}^{-\sum_{i=1}^{n-k}\left(t'_{i}-\ln
x'_{\max,i}/\lambda\right)\left(1-{\rm i}\epsilon\right)/\tD} \nonumber\\\
&& \times {\rm e}^{-\sum_{i=k+1}^{n-1}\ln x'_{i}/(\lambda\tD)}{\rm e}^{-{\rm
i}\epsilon\sum_{i=1}^{k}\ln x'_{i}/(\lambda\tD)} \nonumber \\
&& \times {\rm e}^{-\tau\left(1-{\rm i} n\epsilon \right)} ,
\end{eqnarray}
with $-\ln x'_{\max,i} = \max_{j\in l(i)}\left\lbrace-\ln x'_{j}\right\rbrace$
linked to the duration of the longest encounter stretch contained within $t_i$. 
We have also used $\hat{M}[i,j]$ defined after \eref{mix1} and dropped the
explicit $i$ dependence of $l,L$ and $m$ above. Again we obtain the
Ehrenfest-time dependence predicted by \eref{eq5}.

\begin{figure}
\begin{center}
\includegraphics[width=0.65\columnwidth]{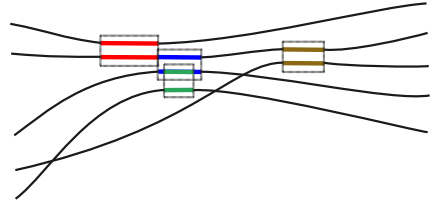}
\caption{(Color online) An encounter diagram containing also orbits with more
than two stretches. In contrast to Fig.\ \ref{fig8} encounter stretches here are
allowed to be contained fully inside others. The encounters are marked by boxes
(shown in different colors).}
\label{fig9a}
\end{center}
\end{figure}

As in the case of the ladder diagram above, we can also have the possibility of
some encounter stretches being contained in larger encounter stretches and some
separated from those larger encounters, see Fig.\ \ref{fig9a}
 for an example of a possible diagram.   This just implies that some of the
$t_i$ integrals have to be treated as was done in the case of the configuration
shown in Fig.\ \ref{fig5}, and the Ehrenfest-time dependence predicted by
\eref{eq5} also follows in this case.

\subsubsection{Touching the lead}

When the encounters are allowed to enter the lead we again have to consider
times representing how far each encounter has moved into the lead (actually how
much of the encounter remains inside the system).  As for the case treated in
detail for $n=3$ it is only the upper limit (namely the full encounter time) of
these time integrals that have the necessary encounter time dependence to
contribute in the semiclassical limit.  The reasoning for $n=3$ can then be
carried over directly to the more general cases as the upper limits of these
integrations yield contributions that are (up to constant factors) the same as
the ones obtained when the encounters are inside the system.  We thus obtain the
same Ehrenfest-time dependence from encounters moved into the leads.

\subsection{Summary}

The separate diagrams considered in the RMT-type semiclassical treatment
\cite{Kui} can be created from the original collapse of trajectories onto each
other and by sliding the individual encounters together or into the leads.  The
Ehrenfest-time treatment however suggests treating all of these possibilities
instead as part of continuous families.  What we have shown above in this
section is that, if we partition this family in a particular way, for any
partition we can find a suitable set of coordinates that allows us to transform
the semiclassical contribution so that we can extract the overall Ehrenfest-time
dependence.  Though the exact details of this transformation depend on the
structures of the partition, the algorithmic routines described above all lead
to the same Ehrenfest-time dependence.  Each partition and hence family then has
the factor ${\rm e}^{-\tau\left(1-{\rm i} n\epsilon \right)}$ and no other
Ehrenfest-time or $\hbar$ dependence.  As we know that we must recover the
RMT-type result $C(\epsilon,n)$ in \eref{eq5} when $\tE=0$ (since we treat the
same diagrams) with no further Ehrenfest-time dependence, we then obtain the
full result in \eref{eq5} and hence provide a semiclassical justification of the
effective RMT ansatz.

\section{Trajectories always correlated} \label{part2}

\begin{figure}
\begin{center}
\includegraphics[width=0.5\columnwidth]{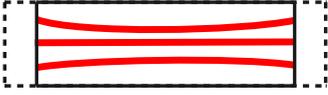}
\caption{(Color online) Band of $n=3$ correlated trajectories. The length of the
orbits is marked by a box; the duration of the encounter $t_{\rm
enc}=1/\lambda\ln
\left(c^2/{\max}_i\left|s_i\right|{\max}_j\left|u_j\right|\right)$ is marked by
a dotted box.} 
\label{fig10}
\end{center}
\end{figure}

In this section we determine the so called classical contribution in \eref{eq6}.
To obtain this contribution $C_{2}(\epsilon,\tau,n)$ semiclassically we consider
a band of $n$ trajectories that are correlated (inside the same encounter) for
their entire duration between entering and leaving the system as in Fig.\
\ref{fig10}.  This implies that all the trajectories have the same length $t$
and that the maximum of the differences $s_i,u_i$ between their stable and
unstable coordinates lies below the constant $c$ (related to the Ehrenfest
time).  For the case $n=2$ this configuration was first considered in
\ocite{Whi} and then extended to $n=3$ in \ocite{Bro}.  For our calculation we
follow \ocite{Bro} and place a PSS at a distance $t_1$ from the left end of the
trajectories while the remaining time on the right of the section is denoted by
$t_2=t-t_1$.   The semiclassical contribution $C_{2}(\epsilon,\tau,n)$ can be
written as
\begin{eqnarray}
\label{allcorr}
C_{2}(\epsilon,\tau,n)&=&\frac{1}{\tD}\int_0^\infty {\rm
d}t_{1}\int_0^{\infty}{\rm d}t_2\frac{{\rm e}^{-t\left(1-{\rm
i}n\epsilon\right)/\tD}}{(2\pi\hbar)^{n-1}\left(t_1+t_2\right)}\nonumber\\ &&
\times\int_{|s_i|\leq c {\rm e}^{-\lambda t_1}}\hspace{-2em}{\rm
d}s^{n-1}\int_{|u_i|\leq c {\rm e}^{-\lambda t_2}}\hspace{-2em}{\rm
d}u^{n-1}{\rm e}^{\left({\rm i}/\hbar\right)\Delta S} , \nonumber \\
&&
\end{eqnarray}
where we only include one traversal of the band in the survival probability and
the restrictions on the $s$ and $u$ integrals ensure that the band always
remains together under the exponential divergence of the trajectories due to the
chaotic dynamics.  Performing an integral over $t_1-t_2$ and the $u_i$
integrals gives
\begin{eqnarray}
\label{eq35}
C_{2}(\epsilon,\tau,n)&=&\frac{4^{n-1}}{\tD}\int_0^\infty {\rm d}t\, \frac{{\rm
e}^{-t\left(1-{\rm i}n\epsilon\right)/\tD}}{(2\pi\hbar)^{n-1}}\\ 
&&\times\int_0^{{\rm e}^{-\lambda t}}{\rm
d}x^{n-1}\prod_{i=1}^{n-1}\frac{\hbar}{x_i}\sin\left(\frac{c^{2}x_i}{\hbar}
\right) , \nonumber
\end{eqnarray}
where $x_{i}={\rm e}^{-\lambda t_2}s_{i}/c$.  Using that in the semiclassical
limit
\begin{equation}
\label{eq36}
\int_0^{{\rm e}^{-\lambda t}}{\rm
d}x\frac{\hbar}{x}\sin\left(\frac{c^{2}x}{\hbar}\right)=\frac{\pi\hbar}{2}
\Theta\left(\tE-t\right) ,
\end{equation}
with the Heaviside theta function $\Theta(x)$, we finally obtain
\begin{equation}
C_{2}(\epsilon,\tau,n) =  \frac{1-{\rm e}^{-\tau(1-{\rm i}n\epsilon)}}{1-{\rm
i}n\epsilon} ,
\end{equation}
proving the Ehrenfest-time dependence of the $C_{2}(\epsilon,\tau,n)$ in
\eref{eq6}.

\section{Mixed Terms}\label{mixing}

\begin{figure}
\begin{center}
\includegraphics[width=0.5\columnwidth]{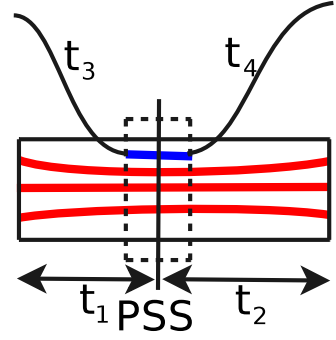}
\caption{(Color online) An example of a band of three trajectories that possesses
an encounter with another trajectory. The band is marked by a thicker box (red
stretches), the encounter of the other trajectory with the band by a dotted
box (blue stretch) and the links by thin (black) lines. The link durations are
denoted by $t_3$ and $t_4$, and the durations of the band before and after the PSS
by $t_1$ and $t_2$.} 
\label{fig11}
\end{center}
\end{figure}

Finally we want to consider possible correlations between trajectory structures
giving the RMT-type contribution and those giving the classical part, i.e.\
contributions from correlations between bands of trajectories (that are always
correlated with each other) and trajectories that are only correlated with each
other during encounters.   In particular we want to show that diagrams that have
a correlated band that has any encounter with other trajectory structures (with
encounters) give no contribution in the semiclassical limit. This, once
generalized, then excludes the existence of mixed terms in \eref{eq4} so that
\eref{eq4} is complete.   First we consider the case in which $n-1$ of the
trajectories form a correlated band with the remaining trajectory meeting the
band in an encounter inside the system as depicted in Fig.\ \ref{fig11}.  This
contribution $C^{\ref{fig11}}(\epsilon,\tau,n)$ to the correlation function
$C(\epsilon,\tau,n)$ can be written by treating the correlated band as before
and introducing the times $t_3$ and $t_4$ to represent the durations of the
parts of the trajectory that encounter the band on the left and on the right of
the PSS.  It reads
\begin{eqnarray}
\label{eq38}
 C^{\ref{fig11}}(\epsilon,\tau,n)&=&\frac{1}{\tD^{3}}\int_0^\infty
\prod_{i=1}^{4}{\rm d}t_{i}\frac{{\rm e}^{-\sum_{i=1}^{4}t_{i}\left(1-{\rm
i}\epsilon\right)/\tD}}{(2\pi\hbar)^{n-1}} \hspace{2em} \nonumber\\
 && \times\prod_{i=1}^{n-2}\left[\int_{|s_i|\leq c {\rm e}^{-\lambda
t_1}}\hspace{-2em}{\rm d}s^{n-2}\int_{|u_i|\leq c {\rm e}^{-\lambda
t_2}}\hspace{-2em}{\rm d}u^{n-2}\right]\nonumber \\
 && \times\int_{c {\rm e}^{-\lambda t_1}<|s|\leq c}\hspace{-1em}{\rm d}s\int_{c
{\rm e}^{-\lambda t_2}<|u|\leq c}\hspace{-1em}{\rm d}u  \nonumber \\
&& \times \frac{{\rm e}^{\left({\rm i}/\hbar\right)\Delta S}{\rm e}^{{\rm
i}\epsilon \left(t_{\rm enc}+(n-2)(t_1+t_2)\right)/\tD}}{\left(t_1+t_2\right)} ,
\end{eqnarray} 
where $t_{\rm enc}$ is the time during which the remaining trajectory encounters
the band. In this case no overcounting factor $1/t_{\rm enc}$ occurs as the
limits of the $s,u$-integrals in the third line are chosen such that the time
integral accounting for all possible positions of the single encounter stretch
with respect to the band cancels this factor.  Choosing different limits for the
$s,u$-integrals we could replace the current overcounting factor of
$1/\left(t_1+t_2\right)$ in \eref{eq38} again by the usual $1/t_{\rm enc}$ as in
Sec.\ \ref{part1}.

The integrals over the $(n-2)$ $s,u$-coordinates within the band are performed
in the same manner as in the last section in \eref{eq36} and lead to the same
Heaviside function $\Theta(\tE-t)$.
For the integrals in the third line in \eref{eq38} over the differences $s,u$
between the coordinates of a band trajectory and the trajectory encountering it
we obtain 
\begin{eqnarray}
\label{eq38b}
&&\int_{c {\rm e}^{-\lambda t_1}<|s|\leq c}\hspace{-1em}{\rm d}s\int_{c {\rm
e}^{-\lambda t_2}<|u|\leq c}\hspace{-1em}{\rm d}u\:{\rm e}^{\left({\rm
i}/\hbar\right)su}{\rm e}^{{\rm i}\epsilon t_{\rm
enc}/\tD}\nonumber\\&\approx&\int_{{\rm e}^{-\lambda t}}^1{\rm
d}x\frac{4\hbar}{x}\sin\left(\frac{c^2x}{\hbar}\right){\rm e}^{{\rm i}\epsilon
\tau}
\nonumber\\&=&\int_{{\rm e}^{-\lambda \left( t-\tE\right) }}^{{\rm
e}^{\lambda\tE}}{\rm d}x'\frac{4\hbar}{x}\sin x' {\rm e}^{{\rm i}\epsilon
\tau}\nonumber \\
&=&{2\pi\hbar}\Theta\left(t-\tE\right){\rm e}^{{\rm i}\epsilon \tau} ,
\end{eqnarray}
with $x={\rm e}^{-\lambda t_2}s/c$.  This Heaviside function is opposite to the
one from the band, so that the total contribution vanishes due to the opposing
restrictions of the theta functions. Note that when evaluating the
Ehrenfest-time dependence of the RMT contributions in Sec.\ \ref{part1} we
chose a different way for performing the integrals. In Sec.\ \ref{part1} we
performed restricted time integrals first and unrestricted $x'$ integrals
afterwards. The different order of calculating these integrals leads to the
theta functions which make it clear that mixed terms like in Fig.\ \ref{fig11}
vanish; this is why we use this ordering here. If we move more trajectories
from the band (composed of at least two trajectories) to the trajectory
structure with encounters we still obtain these opposing Heaviside functions and
hence no contribution.

A similar reasoning can be applied if the encounter of a trajectory (or part of
a trajectory structure) with a band does not happen inside the system but enters
the lead at the beginning or the end. In this case we obtain an additional time
integral with respect to the time of the encounter that remains inside the
system but, as the $s,u$-integrals still yield the same Heaviside functions,
this contribution also vanishes. Note that if we move both ends of the encounter
into the leads then the encountering trajectory can be considered as part of the
band and treated as above or in Sec.\ \ref{part2}.

The reasoning in this section applies to an arbitrary number of bands of
correlated trajectories connected by trajectories that are only correlated in
encounters.  Therefore all such mixed terms vanish.

\section{Implications for transport and scattering}

\subsection{Moments of transmission}

Up to now we have concentrated on energy dependent correlation functions involving
the whole scattering matrix. Because we use the same semiclassical diagrams our
result can be applied directly to dc-transport properties of chaotic systems
such as the moments of the transmission or reflection eigenvalues.  Assuming the
system has two scattering leads and taking just the transmission subblock $t$ of
the scattering matrix connecting the $N_1$ channels in lead 1 to the $N_2$
channels in lead 2 (without an energy difference) the moments of the
transmission eigenvalues can be written as
\begin{equation}
M(\tau,n) = \frac{1}{N}{\rm Tr} \left[t t^{\dagger}\right]^{n},
\end{equation}
where $N=N_1+N_2$ is the total number of channels. Within the
Landauer-B\"uttiker\cite{Lan} approach to quantum transport the moments carry
information about statistical properties  of the transmission in the phase
coherent regime, the counting statistics\cite{Naz}. For example the first moment
characterizes the average conductance $G\propto M(\tau,1)$ and the second one
the power of the shot noise $P\propto M(\tau,1)-M(\tau,2)$. Using the
semiclassical results from this paper we can simply write
\begin{equation}
\label{eq40}
M(\tau,n) = M(n){\rm e}^{-\tau} + \frac{N_1 N_2}{N^2}\left(1-{\rm
e}^{-\tau}\right) ,
\end{equation}
where we have included the probability $N_1/N$ of starting in lead 1 in the
moments. Equation (\ref{eq40}) can be generalized to ac-transport considered in
\ocite{Petit} by including in the latter equation the $\epsilon$-dependent
factors given in (\ref{eq5},\ref{eq6}). The result in \eref{eq40} again splits
into two parts with the first involving the semiclassical moments $M(n)$
calculated in \ocite{Ber}, which in turn lead to the random matrix probability
distribution \cite{bb96,novaes07}.

The second term in \eref{eq40} leads to the classical Bernoulli distribution
where the transmission amplitude $T$ is 1 with probability $N_2/N$ and 0
otherwise (i.e.\ $N_1/N$).  The Ehrenfest time then provides a smooth
interpolation between these two distributions giving a weight ${\rm e}^{-\tau}$
to the RMT one and the remaining weight $(1-{\rm e}^{-\tau})$ to the classical
one.  A similar formula and result follows for the moments and probability
distribution of the reflection eigenvalues whose zero Ehrenfest time
contributions can be simply derived from the treatment in \ocite{Ber}.

\subsection{Moments of delay times}

Taking the full correlation functions $C(\epsilon,\tau,n)$ it is possible to
obtain not only the Ehrenfest-time dependence of the density of states of
chaotic Andreev systems, covered in detail in \ocite{Kui}, but also the moments
and distribution of the Wigner delay times.  We start with the Wigner-Smith
matrix \cite{Wig}
\begin{equation}
  \label{eq41}
  Q = \frac{\hbar}{\rm i} S^{\dagger}(E) \frac{{\rm d} S(E)}{{\rm d} E} ,
\end{equation}
which can be shown to be Hermitian by using the unitarity of the scattering
matrix. Because of this unitarity $Q$ can also be written as
\begin{equation}
  Q =  \frac{\tD}{{\rm i}}\frac{{\rm d}}{{\rm d}\epsilon}\left[
    S^{\dagger}\left(-\frac{\epsilon\hbar}{2\tD}\right)
    S\left(+\frac{\epsilon\hbar}{2\tD}\right)\right]\Big\vert_{\epsilon=0} ,
\end{equation}
where the scattering matrix energy differences are measured with respect to
the energy $E$.
The delay times are simply the eigenvalues of $Q$ so their moments are
\begin{equation}
  m(\tau,n) = \frac{1}{N}{\rm Tr} \left[Q\right]^{n} .
\end{equation}
Using the relation
\begin{equation}
\label{eq44}
  \frac{1}{n!}\frac{{\rm d}^{n}}{{\rm d}\epsilon^{n}}
  \left[f(\epsilon)-f(0)\right]^{n}\Big\vert_{\epsilon=0}
  = \left[f'(0)\right]^{n} ,
\end{equation}
the moments of the delay times are \cite{Ber2}
\begin{eqnarray}
  \label{eq45}
  m(\tau,n) &=& \frac{\tD^{n}}{{\rm i}^{n}n!N}
  \frac{{\rm d}^{n}}{{\rm d}\epsilon^{n}}{\rm Tr} \\
&& \left[
    S^{\dagger}\left(-\frac{\epsilon\hbar}{2\tD}\right)
    S\left(+\frac{\epsilon\hbar}{2\tD}\right)-I\right]^n\Big\vert_{\epsilon=0} .
\nonumber
\end{eqnarray}
Expanding \eref{eq45}, the moments can be expressed as
\begin{equation}
  \label{eq46}
  m(\tau,n) = \frac{\tD^{n}}{{\rm i}^{n}n!}
  \frac{{\rm d}^{n}}{{\rm d}\epsilon^{n}} \sum_{k=1}^{n}
  \left(-1\right)^{n-k} \left(\begin{array}{c}n\\ k \end{array}\right)
  C(\epsilon,\tau,k)\Big\vert_{\epsilon=0} ,
\end{equation}
in terms of the correlation functions calculated before.  As this is additive we
can look at the two parts of the Ehrenfest-time dependent results in \eref{eq4}
separately.

For the first part of the $C_1(\epsilon,\tau,k)$ however, it is possible to put
our Ehrenfest dependence into the framework of \ocite{Ber2} where the moments
and the probability distribution $\rho(\tW)$ of the Wigner delay times $\tW$
were calculated (without the Ehrenfest time).  We can actually obtain the result
in a simple way.  Using \eref{eq44} again we can see that since
\begin{equation}
  \label{eq47}
  Q - \tE I =  \frac{\tD}{{\rm i}}\frac{{\rm d}}{{\rm d}\epsilon}\left[
    S^{\dagger}\left(-\frac{\epsilon\hbar}{2\tD}\right)
    S\left(+\frac{\epsilon\hbar}{2\tD}\right){\rm e}^{-{\rm
i}\epsilon\tau}\right]\Big\vert_{\epsilon=0} ,
\end{equation}
we have
\begin{eqnarray}
  \label{eq48}
  \frac{1}{N}{\rm Tr} \left[Q-\tE I\right]^{n} & = & \frac{\tD^{n}}{{\rm
i}^{n}n!}
  \frac{{\rm d}^{n}}{{\rm d}\epsilon^{n}} \sum_{k=1}^{n} \\
& & \left(-1\right)^{n-k} \left(\begin{array}{c}n\\ k \end{array}\right)
  C(\epsilon,\tau,k){\rm e}^{-{\rm i}k\epsilon\tau}\Big\vert_{\epsilon=0} .
\nonumber
\end{eqnarray}
Plugging in our result for $C_1(\epsilon,\tau,k)$ from \eref{eq5} the energy
dependent exponentials cancel so, apart from the damping factor ${\rm
e}^{-\tau}$, we just have the moments without any Ehrenfest-time dependence,
leading \cite{Ber2} to the RMT result \cite{Bro99}.  Of course on the left hand
side of \eref{eq48} we have a simple translation by the Ehrenfest time, meaning
that the translated probability distribution is the same as the RMT-one
(damped).  For the full Ehrenfest-time dependent distribution we simply
translate back again and have
\begin{eqnarray}
\label{eq49}
\rho_1(\tW)&=&\frac{\sqrt{\left(\tau_{+}-\tW\right)
\left(\tW-\tau_{-}\right)}}{2\pi(\tW-\tE)^{2}} {\rm e}^{-\tau},\quad
\tau_{-}<\tW<\tau_{+} \nonumber \\
\tau_{\pm}&=&(3\pm\sqrt{8})\tD+\tE .
\end{eqnarray}

For the second `classical' contribution in \eref{eq4} we first take the simplest
part of the contribution
\begin{equation}
C_{2}^{(1)}(\epsilon,\tau,k)=\frac{1}{1-{\rm i}k\epsilon} ,
\end{equation}
and substitute into \eref{eq46} obtaining
\begin{equation}
  \label{eq51}
  m_{2}^{(1)}(\tau,n) = \tD^{n}\sum_{k=1}^{n}
  \left(-1\right)^{n-k} \left(\begin{array}{c}n\\ k \end{array}\right) k^{n} =
\tD^{n} n!
\end{equation}
These moments clearly come from an exponential distribution, so that in the
limit $\tau\to\infty$ we recover, for the probability distribution
$\rho_{2}^{(1)}(\tW)$, the classical exponential decay of trajectories inside
the system
\begin{equation}
\rho_{2}^{(1)}(\tW)=\frac{1}{\tD}{\rm e}^{-\tW/\tD}, \quad \tW>0.
\end{equation}

For the remaining contribution of the second part $\rho_{2}^{(2)}(\tW)$, we have
the damping factor ${\rm e}^{-\tau}$, and the energy dependent phase again just
leads to a shift in the exponential distribution so this contribution starts at
$\tE$, thus yielding 
\begin{equation}
\rho_{2}^{(2)}(\tW)=-\frac{1}{\tD}{\rm e}^{-\tW/\tD}, \quad \tW>\tE.
\end{equation}
The minus sign however means we truncate the previous exponential at $\tE$, so
the total second contribution to the probability distribution is
\begin{equation}
\rho_2(\tW)=\frac{1}{\tD}{\rm e}^{-\tW/\tD}, \quad 0<\tW<\tE ,
\end{equation}
and 0 elsewhere. Since this contribution to the delay time probability
distribution is considered to be made up of the two distributions
$\rho_{1}^{(1)}(\tW)$ and $\rho_{1}^{(2)}(\tW)$ and as the shifted and damped
one $\rho_{1}^{(2)}(\tW)$ has the same mean ($\tD+\tE$) and weight as the
shifted and damped RMT distribution $\rho_{1}(\tW)$ but a minus sign, it is
clear that the average time delay stays at $\tD$ (i.e.\ from the untruncated
exponential distribution) and is unaffected by the Ehrenfest time.  In fact this
is an example of the general relation \cite{Lyu}, derived from the unitarity of
the scattering matrix, in which the mean time delay depends just on the average
spacing of the resonant levels of the scattering system and the number of
scattering channels, i.e.\ it should have no Ehrenfest time or other dependence.

\begin{figure}
\begin{center}
\includegraphics[width=0.8\columnwidth]{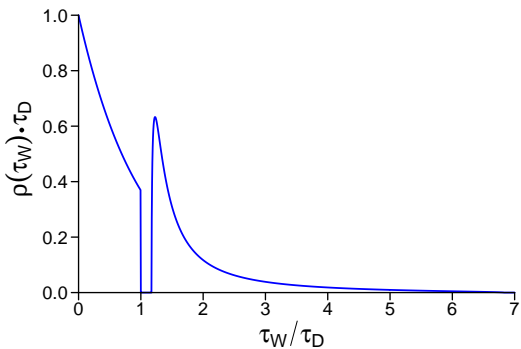}
\caption{(Color online) The probability density of the Wigner delay times for
$\tau=1$.}
\label{fig12}
\end{center}
\end{figure}

This shape of the distribution however is significantly affected by the
Ehrenfest time, and as an example we plot the complete probability density of
the Wigner delay times for $\tau=1$ in Fig.\ \ref{fig12}.  There we can see the
exponential decay truncated at $\tE=\tD$ before a hard gap separates it from the
damped RMT distribution which is shifted to the right by $\tE$.  As the RMT
distribution in \eref{eq49} starts at $\tau_{-}$ which is $(3-\sqrt{8})\tD$
above the Ehrenfest time, the hard gap is a constant $0.172\tD$ wide.  The RMT
distribution is continuous (unlike the truncation of the exponential at $\tE$)
and peaks very quickly ($0.056\tD$) after the gap at around $1.72{\rm
e}^{-\tau}/\tD$.  Related to the semicircle distribution, it likewise has an
upper bound ($\tau_{+}$), which interestingly can be connected \cite{Lib} to the
excitation gap in the density of states of Andreev billiards.  By expanding, for
small energies, the determinantal equation that governs this excitation gap in
terms of the Wigner-Smith matrix $Q$, \ocite{Lib} showed that the gap's width is
approximately the inverse of the maximum delay time.  There the maximum was
increased by disorder scattering, leading to a decrease in the Andreev gap,
while here the maximum delay time is increased by the Ehrenfest time as the RMT
distribution is correspondingly shifted right.  A simultaneously decreasing
Andreev gap fits with the effective RMT \cite{Goorden,Ben} and semiclassical
\cite{Kui} treatments.  However the connection between the secondary gaps, which
appear in the density of states of Andreev billiards \cite{Kui} as a consequence
of the Ehrenfest dependence in \erefss{eq4}{eq5}{eq6} shown in this paper, and
the gap in the density of the delay times is not so clear.

\section{Conclusions}

In this paper we have shown how to treat the effect of the Ehrenfest time on
correlation functions of arbitrarily many pairs of scattering matrices.  In our
semiclassical approach we extended and combined the zero Ehrenfest time approach
\cite{Kui} (which leads to the RMT result) and the $n=3$ Ehrenfest time approach
\cite{Bro} and showed how the results of the effective RMT ansatz can be
obtained.  The different contributions are described by simple diagrams and
following an innovative way of partitioning these diagrams we implemented an
algorithmic procedure that allows one to easily obtain the Ehrenfest-time
dependence.  Interestingly this always led to the same factor (which can be
traced back to the survival probability only depending on one traversal of each
encounter) so that the RMT-type expression is simply modified by the Ehrenfest
time by the additional factor ${\rm e}^{-\tau(1-{\rm i}n\epsilon)}$.  This is in
line with the effective RMT result, but as our result is derived just from the
underlying chaotic dynamics of the system we can justify for this situation the
use of effective RMT which instead conjectures the Ehrenfest-time dependence.

As the semiclassical framework is based on the underlying classical dynamics we
can equally well move away from the RMT arena and obtain the `classical'
contribution to the correlation functions.  This can be seen to come from bands
of trajectories that remain correlated with each other for the entire duration
of their stay inside the system.  Furthermore the fact that no mixed (between
the RMT-type and classical-type) terms arise is simply due to their opposing
classical restrictions.  This lack of mixed terms as well as the classical
contribution were previously shown to be more generally due to the preservation
of volume under the dynamical evolution and the separation of phase-space into
two essentially independent subsystems \cite{Jac,Jacquod}.

The separation of the correlation functions into two contributions, which each
have a straightforward dependence on the Ehrenfest time was previously shown to
be responsible for secondary gaps in the density of states of Andreev billiards
\cite{Kui}, but has an equal effect on other transport quantities.  For the
transmission eigenvalues (and their moments) with no energy dependence we just
get a straightforward interpolation between the RMT \cite{Ber} and classical
values.  For the distribution of the Wigner delay times we further see a
truncation of the classical (exponential) distribution and a shift to higher
times of the RMT-type distribution.  Between the two though a hard gap remains.

The method described in this paper allows for the computation of the
Ehrenfest-time dependence of the trace of arbitrarily many scattering matrix
pairs but only to leading order in inverse channel number $1/N$.  The
calculation was only doable because at this order the corresponding
semiclassical diagrams involve no closed loops and have no periodic orbit
encounters (surrounded periodic orbits).  When such surrounded periodic orbits
are involved, for example for the conductance variance \cite{Bro1} or the next
to leading order quantum correction to the transmission, reflection and the
spectral form factor \cite{Wal1}, the relatively simple cancellation mechanism
observed in this paper no longer holds.  But by taking into account all
possibilities for partial correlations within the `fringes' as in
\ocites{Wal1,Bro1} in a systematic way, it should however also be possible to
extend our Ehrenfest-time results to infinite order.

\begin{acknowledgments}
The authors thank Cyril Petitjean and Robert Whitney for helpful discussions and
acknowledge funding through the Deutsche Forschungsgemeinschaft (Forschergruppe
FOR 760) (KR) and the Alexander von Humboldt Foundation (JK).
\end{acknowledgments}

\end{document}